\documentstyle[preprint,aps,epsfig,floats]{revtex}

\def\lpa{\lambda_{p{-}\rm air}}
\def\spa{\sigma_{p{-}\rm air}}
\def\spai{\sigma_{p{-}\rm air}^{\rm inel}}
\def\spae{\sigma_{p{-}\rm air}^{\rm el}}
\def\spaqe{\sigma_{p{-}\rm air}^{q{-}\rm el}}

\catcode`@=11
\def\references{%
\ifpreprintsty
\bigskip\bigskip
\hbox to\hsize{\hss\large \refname\hss}%
\else
\vskip24pt
\hrule width\hsize\relax
\vskip 1.6cm
\fi
\list{\@biblabel{\arabic{enumiv}}}%
{\labelwidth\WidestRefLabelThusFar  \labelsep4pt %
\leftmargin\labelwidth %
\advance\leftmargin\labelsep %
\ifdim\baselinestretch pt>1 pt %
\parsep  4pt\relax %
\else %
\parsep  0pt\relax %
\fi
\itemsep\parsep %
\usecounter{enumiv}%
\let\p@enumiv\@empty
\def\theenumiv{\arabic{enumiv}}%
}%
\let\newblock\relax %
\sloppy\clubpenalty4000\widowpenalty4000
\sfcode`\.=1000\relax
\ifpreprintsty\else\small\fi
}
\catcode`@=12

\begin{document}
\tightenlines

\preprint{\font\fortssbx=cmssbx10 scaled \magstep2
\hbox to \hsize{
\includegraphics{uwlogo.ps}
\hfill$\vcenter{\tighten
                \hbox{\bf  MADPH-99-1132}
		    \hbox{\bf  NUHEP 602} 	
                    \hbox{\bf   BA-99-56}
                    \hbox{\bf hep-ph/9908222}
                \hbox{August 1999}}$}}

\title{\vspace*{.5in}
Predicting Proton-air Cross Sections at $\sqrt s\sim30$~TeV,\\
using Accelerator and Cosmic Ray Data}

\author{M.~M. Block$^1$, Francis Halzen$^2$ and Todor Stanev$^3$}

\address{$^1$Department of Physics and Astronomy, Northwestern University,
Evanston, IL 60208\\
$^2$Physics Department, University of Wisconsin, Madison, WI 53706\\
$^3$Bartol Research Institute, University of Delaware, Newark, DE 19716}

\maketitle

\begin{abstract}
\vskip-.35in
 We use the high energy predictions of a QCD-inspired parameterization
 of all accelerator data on forward proton-proton and antiproton-proton
 scattering amplitudes, along with Glauber theory, to predict proton--air
 cross sections at energies near $\sqrt s \approx$ 30 TeV.   The
 parameterization of the proton-proton cross section incorporates
 analyticity and unitarity, and demands that the asymptotic proton is a
 black disk of soft partons. By comparing with the p-air cosmic ray
 measurements, our  analysis results in a constraint on the inclusive
 particle production cross section.
\end{abstract}

\thispagestyle{empty}
\vskip.65in

 Cosmic ray experiments measure the penetration in the atmosphere
 of particles with energies in excess of those accelerated by existing
 machines---interestingly, their energy range covers the energy of the
 Large Hadron Collider (LHC) and extends beyond it. However, extracting
 proton--proton cross sections from cosmic ray observations is far from
 straightforward~\cite{gaisser}. By a variety of experimental techniques,
 cosmic ray experiments map the atmospheric depth at which cosmic ray
 initiated showers develop.
 The measured shower attenuation length ($\Lambda_m$) is not only
 sensitive to the interaction length of the protons in the atmosphere
 ($\lpa$), with
\begin{equation}
\Lambda_m = k \lpa = k { 14.5 m_p \over \spai} \,,  \label{eq:Lambda_m}
\end{equation}
 but also depends on the rate at which the energy of the primary proton
 is dissipated into electromagnetic shower energy observed in the
 experiment. The latter effect is parameterized in Eq.\,(\ref{eq:Lambda_m})
 by the parameter $k$; $m_p$ is the proton mass and $\spai$ the inelastic
 proton-air cross section. The value of $k$ depends on the inclusive
 particle production cross section in nucleon and meson interactions
 on the light nuclear target of the atmosphere and its energy dependence.
 We here ignored the fact that particles in the cosmic  ray "beam" may be
 nuclei, not just protons. Experiments allow for this by omitting
 from their analysis showers which dissipate their energy high in the
 atmosphere, a signature that the initial energy is distributed over the
 constituents of a nucleus.

\begin{figure}
\begin{center}
\mbox{\epsfig{file=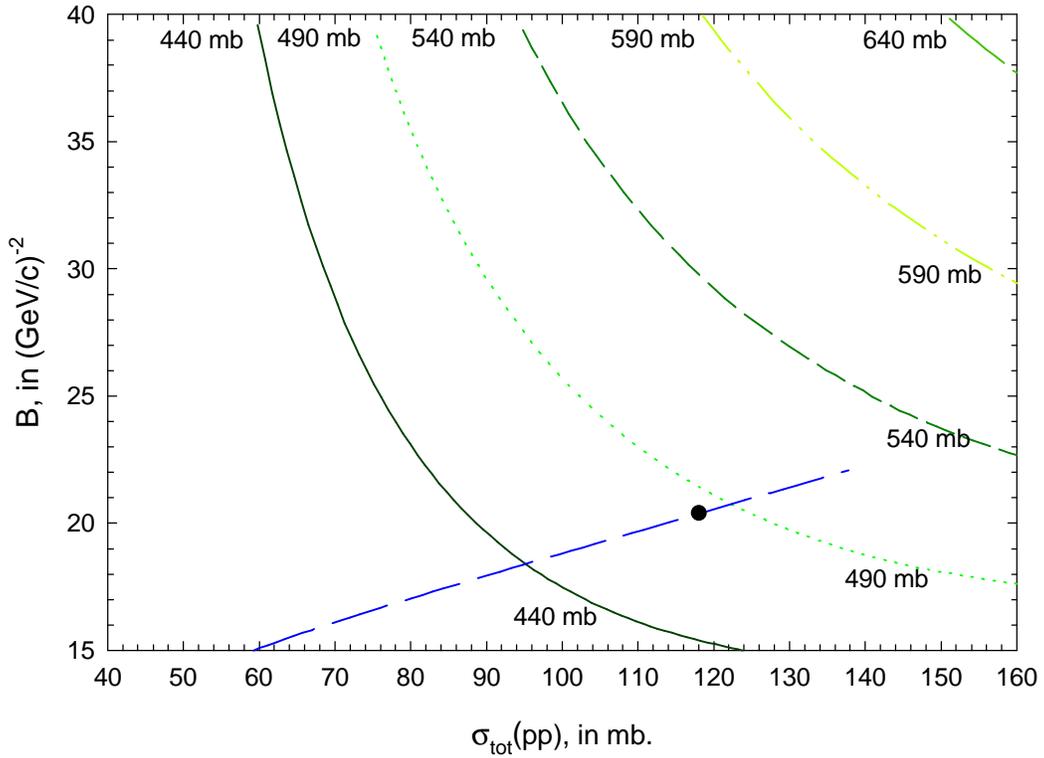%
              ,width=6in,bbllx=85pt,bblly=350pt,bburx=520pt,bbury=675pt,clip=%
}}
\end{center}
\caption[]{  $B$ dependence on the pp total cross section $\sigma_{pp}$. The
 five curves are lines  of constant  $\spai$,  of 440, 490, 540, 590 and
 640 mb---the central value is the published Fly's Eye value, and the others
 are $\pm 1\sigma$ and $\pm 2\sigma$. The dashed curve is a plot of our
 QCD-inspired fit of $B$ against $\sigma_{pp}$.  The dot is our value for
 $\sqrt s=30$ TeV, the Fly's Eye energy.}
\label{fig:p-air}
\end{figure}

 The extraction of the pp cross section from the cosmic ray data is a two
 step process. First, one calculates the $p$-air total cross section from
 the measured inelastic cross section
\begin{equation}
\spai = \spa - \spae - \spaqe \,.  \label{eq:spa}
\end{equation}
 Next, the Glauber method\cite{yodh} is used to transform the measured
 value of $\spai$ into a proton--proton total cross section $\sigma_{pp}$;
 all the necessary steps are calculable in the theory. In Eq.\,(\ref{eq:spa})
 the measured cross section for particle production is supplemented with
$\spae$ and $\spaqe$, 
 the elastic and quasi-elastic cross section, respectively, as calculated by the
 Glauber theory, to obtain the total cross section $\spa$. The subsequent
 relation between $\spai$ and $\sigma_{pp}$ involves the slope of the
 forward scattering amplitude for elastic $pp$ scattering, ${d\sigma_{pp}^{\rm el}\over dt}$,
\begin{equation}
B = \left[ {d\over dt} \left(\ln{d\sigma_{pp}^{\rm el}\over dt}\right)
 \right]_{t=0} \,,
\end{equation}
 and is shown in Fig.\,\ref{fig:p-air}, which plots $B$ against
 $\sigma_{pp}$, for 5 curves of different values of $\spai$.
 This summarizes the reduction procedure
 from $\spai$ to $\sigma_{pp}$~\cite{gaisser}.
 Also plotted in Fig.\,\ref{fig:p-air} is a curve of $B$ {\em vs.}
 $\sigma_{pp}$ which will be discussed later.
	
\begin{figure} 
\begin{center}
\mbox{\epsfig{file=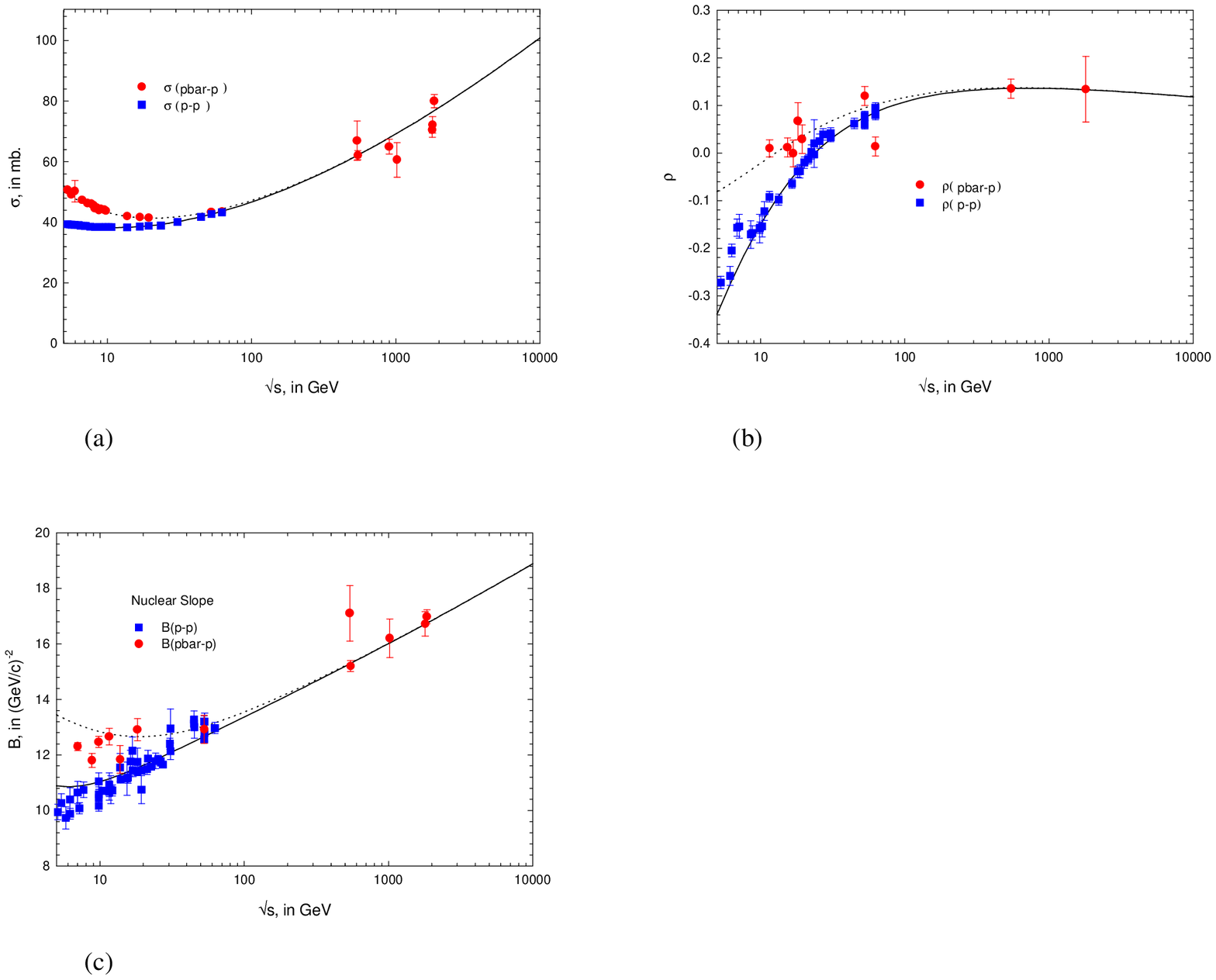%
            ,width=6.7in,bbllx=12pt,bblly=208pt,bburx=600pt,bbury=700pt,clip=%
}}
\end{center}
\caption[]{  The simultaneous QCD-inspired fit of total cross section
 $\sigma_{pp}$, $\rho$ and $B$ {\em vs.} $\sqrt s$, in GeV, for pp
 (squares) and $\bar {\rm   p}$p (circles) accelerator data:
 (a) $\sigma_{pp}$, in mb,\ \  (b) $\rho$,\ \  (c) Nuclear slope $B$,
 in GeV$^{-2}$}
\label{fig:ppcurves}
\end{figure}

 A significant drawback of the method is that one needs a model of
 proton--air interactions to complete the loop between the measured
 attenuation length $\Lambda_m$ and the cross section $\spai$,
 {\em i.e.,} the value of $k$ in Eq. (\ref{eq:Lambda_m}). A proposal
 to minimize the impact of theory is the topic of this letter. We
 have constructed a QCD-inspired parameterization of the forward
 proton--proton and proton--antiproton scattering amplitudes\cite{block}
 which is analytic, unitary and fits all data of $\sigma_{\rm tot}$, $B$
 and $\rho$, the ratio of the real-to-imaginary part of the forward
 scattering amplitude;  see Fig.\,\ref{fig:ppcurves}. We emphasize
 that all 3 quantities are simultaneously fitted. Using vector meson
 dominance and the additive quark models, it accommodates a wealth of
 data on photon-proton and photon-photon interactions without the
 introduction of new parameters\cite{eduardo}. Because the model is
 both unitary and analytic, it has high energy predictions that are
 essentially theory--independent.  In particular, it also
 {\em simultaneously} fits $\sigma_{pp}$ and $B$, forcing a relationship
 between the two. Specifically, the $B$ {\em vs.} $\sigma_{pp}$ prediction
 of the model is shown as the dashed curve in Fig.\,\ref{fig:p-air}. The
 dot corresponds to our prediction of $\sigma_{pp}$ and $B$ at
 $\sqrt s$ = 30 TeV. It is seen to be slightly below the curve for
 490 mb, the lower limit of the Fly's Eye measurement, which was
 made at $\sqrt s\approx$ 30 TeV. %
The percentage error in the prediction of $\sigma_{pp}$ at $\sqrt s=30$ TeV is $\approx 3.3$\%, due to the statistical error in the fitting parameters (see references \cite{block},\cite{eduardo}). 
 
\begin{figure}
\begin{center}
\mbox{\epsfig{file=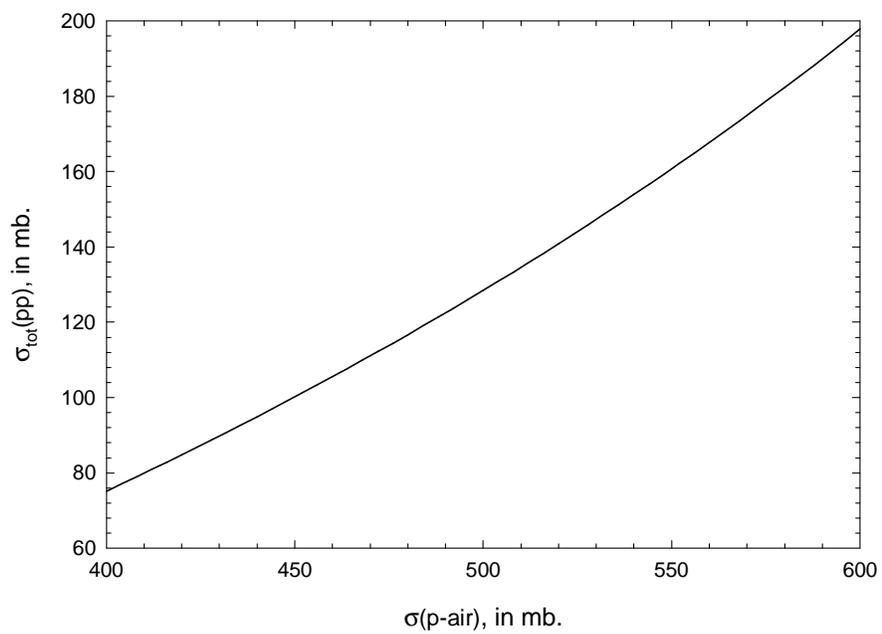%
              ,width=6in,bbllx=55pt,bblly=345pt,bburx=570pt,bbury=680pt,clip=%
}}
\end{center}
\caption[]{A plot of the predicted total pp cross section $\sigma_{pp}$, in mb
 {\em vs.} the measured p-air cross section, $\spai$, in mb.
}
\label{fig:sigpp_p-air}
\end{figure}

In Fig.\,\ref{fig:sigpp_p-air}, we have plotted the values of
 $\sigma_{pp}$ {\em vs.} $\spai$ that are deduced from the
 intersections of the $B$-$\sigma_{pp}$ curve  with the $\spai$
 curves of Fig.\,\ref{fig:p-air}. Figure~\,\ref{fig:sigpp_p-air}
 allows the conversion of the measured $\spai$ to $\sigma_{pp}$ . %
The percentage error in $\spai$ is $\approx 1.9$ \% near $\spai = 450 $mb, due to the error 
in $\sigma_{pp}$ from the model parameter uncertainties.  

\begin{figure}
\begin{center}
\mbox{\epsfig{file=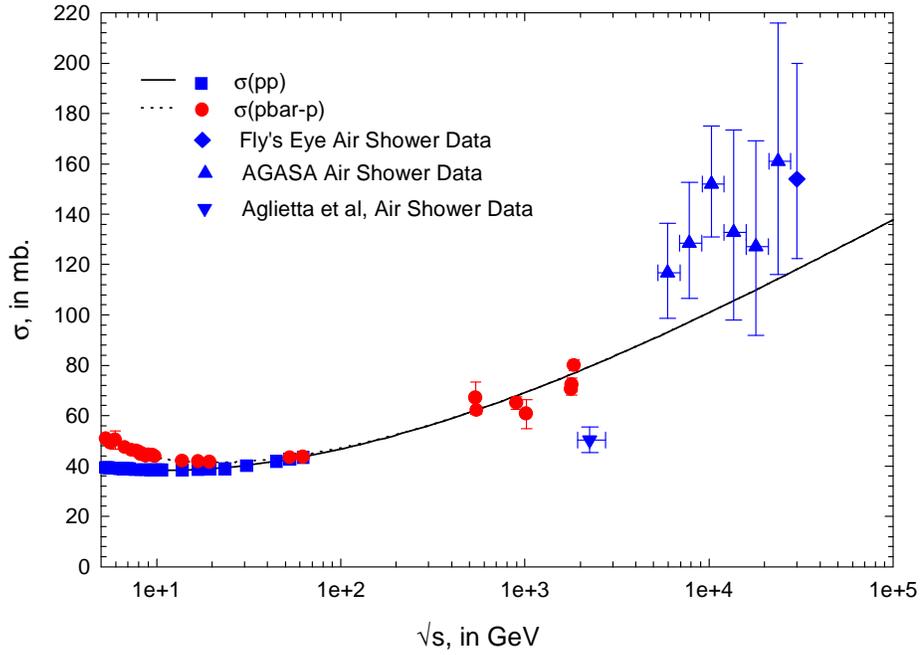%
              ,width=6in,bbllx=85pt,bblly=340pt,bburx=570pt,bbury=680pt,clip=%
}}
\end{center}
\caption[]{\protect
{  A plot of the QCD-inspired fit of the total nucleon-nucleon cross section
 $\sigma_{pp}$, in mb {\em vs.} $\sqrt s$, in Gev. The cosmic ray data that
 are shown have been converted from $\spai$ to $\sigma_{pp}$ using the
 results of Fig.~\ref{fig:sigpp_p-air}.
}
}
\label{fig:sigtodorpp}
\end{figure}

 Our prediction for the total cross section $\sigma_{pp}$ as a
 function of energy is confronted with all of the accelerator and
 cosmic ray measurements\cite{fly,akeno,eastop} in Fig.\,\ref{fig:sigtodorpp}.
 For inclusion in  Fig.\,\ref{fig:sigtodorpp}, we have  calculated
 the cosmic ray values of $\sigma_{pp}$ from the
 {\em published} experimental values of $\spai$, using the results
 of Fig.\,\ref{fig:sigpp_p-air}. We note the systematic underestimate
 of the cosmic ray points, roughly about the level of one standard
 deviation.

 It is at this point important to recall Eq.\,(\ref{eq:Lambda_m})
 and consider the fact that the extraction of  $\spai$ from the
 measurement of $\Lambda_m$ requires a determination of the parameter
 $k$. The measured depth $X_{\rm max}$ at which a shower reaches
 maximum development in the atmosphere, which is the basis of the
 cross section measurement in Ref.~\cite{fly}, is a combined measure
 of the depth of the first interaction, which is determined by
 the inelastic cross section, and of the subsequent shower development,
 which has to be corrected for. The position of $X_{\rm max}$ also
 directly affects the rate of shower attenuation with atmospheric depth
 which is the alternative procedure for extracting $\spai$.

 The model dependent rate of shower development and its fluctuations
 are the origin of the deviation of $k$ from unity
 in Eq.\,(\ref{eq:Lambda_m}). Its values range from 1.5 for a model
 where the inclusive cross section exhibits Feynman scaling, to 1.1
 for models with large scaling violations\cite{gaisser}. The comparison
 between data and experiment in Fig.\,\ref{fig:sigtodorpp} is further
 confused by the fact  that the AGASA\cite{akeno} and Fly's Eye\cite{fly}
 experiments used different values of $k$ in the analysis of their data,
 {\em i.e.,} AGASA used $k=1.5$ and Fly's Eye used $k=1.6$.

\begin{figure}
\begin{center}
\mbox{\epsfig{file=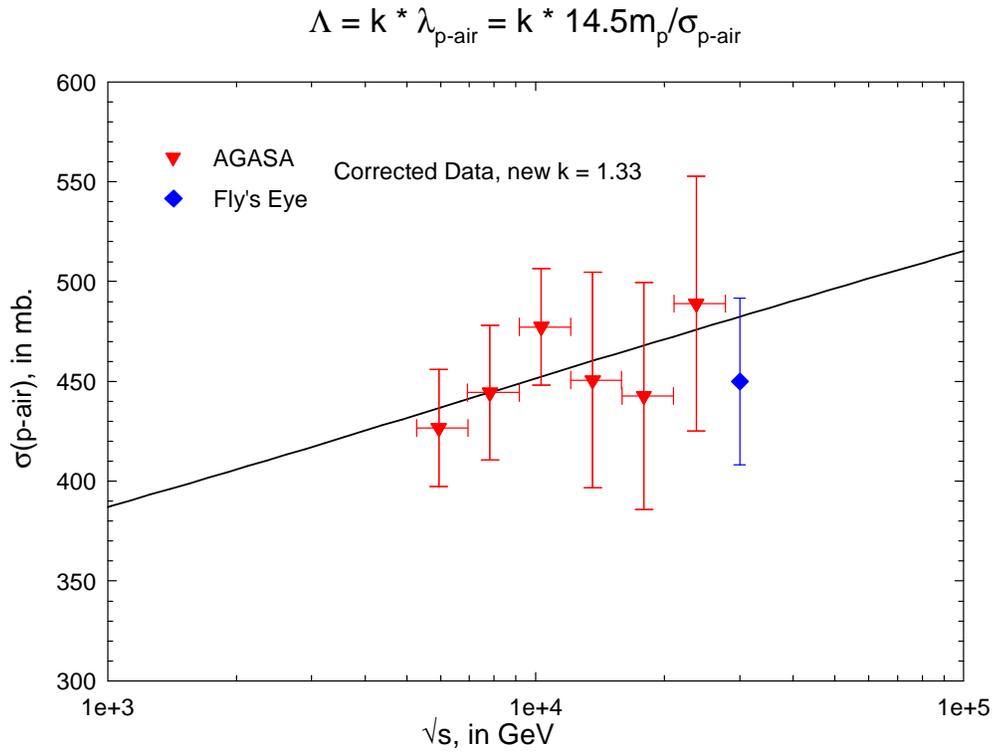%
              ,width=6in,bbllx=68pt,bblly=360pt,bburx=525pt,bbury=720pt,clip=%
}}
\end{center}
\caption[]{\protect
{  A $\chi^2$ fit of the measured AGASA and Fly's Eye data for $\spai$, in mb,
 as a function of the energy, $\sqrt s$, in GeV. The result of the fit for the
 parameter $k$ in Eq. (\ref{eq:Lambda_m}) is $k=1.33\pm0.04$.
}
}
\label{fig:p-aircorrected2}
\end{figure}

 We therefore decided to match the data to our prediction and extract
 a common value for $k$. This neglects the possibility
 that $k$ may show a weak energy dependence over the range measured.
By combining the results of Fig.\,\ref{fig:ppcurves}\,(a) and Fig.\,\ref{fig:sigpp_p-air}, we can plot our prediction of $\spai$ {\em vs.} $\sqrt s$.  To obtain $k$, we leave it as a free parameter and make a $\chi^2$ fit to the rescaled high energy cosmic ray data for $\spai$.
 In Fig.\,\ref{fig:p-aircorrected2} we have replotted the published high energy data for $\spai$, against our prediction of $\spai$ {\em vs.} $\sqrt s$, using the common value of $k=1.33 \pm 0.04\pm 0.026$, obtained from a $\chi^2$ fit.
The error of $0.04$ is the statistical error of the $\chi^2$ fit, whereas the error of $0.026$ is the systematic error due to the error in the prediction of $\spai$.     Clearly,
 we have an excellent fit, with good agreement between AGASA and Fly's Eye. The analysis gives  $\chi^2=1.75$ for 6 degrees of freedom (the low
 $\chi^2$ is probably due to overestimates of experimental errors).  Of course, the improved fit of the cosmic ray data has no effect on the fit to the accelerator data. 

Our result for $k$ is interesting---it is close to the value of $1.2$
 obtained using the SIBYLL simulation\cite{sybill} for inclusive particle
 production. This represents a consistency check in the sense that our
 model for forward scattering amplitudes and SYBILL share the same
 underlying physics. The increase of the total cross section with
 energy to a black disk of soft partons is the shadow of increased
 particle production which is modeled by the production of (mini)-jets
 in QCD. The difference between the $k$ values of 1.20 and 1.33
 could be understood because the experimental measurement integrates
 showers in a relatively wide energy range, which tends to increase
 the value of $k$.

 We predict $108.0\pm3.4$ mb for the total cross section at LHC
 energy (14 TeV), where the error is due to the statistical errors
 of the fitting parameters. In the near term, we look  forward
 to the possibility of repeating this analysis with the higher
 statistics of the HiRes~\cite{HiRes} cosmic ray experiment that
 is currently in progress and the Auger~\cite{Auger} Observatory.

\section*{Acknowledgments}

 This research was supported in part by the U.S.~Department of Energy under
 Grants No.~DA-AC02-76-ER02289 Task B and  No.~DE-FG02-95ER40896 and in part
 by the University of Wisconsin Research Committee with funds granted by the
 Wisconsin Alumni Research Foundation. The research of TS is supported in
 part by the U.S.~Department of Energy under Grant No. DE-FG02-91ER40626.
 One of us (MMB) would like to thank the Aspen Center for Physics for its
 hospitality during the preparation of this manuscript.

\end{document}